\documentclass[aps,prc,reprint,superscriptaddress,showpacs,nofootinbib]{revtex4-1}
\usepackage{lineno,hyperref}
\usepackage{amsmath,amssymb}
\usepackage{braket}
\usepackage{graphicx}
\newcommand{\rp}{\right)}
\newcommand{\lp}{\left(}

\newcommand{\del}{\partial}
\newcommand{\tilZ}{\tilde{Z}}
\newcommand{\calJ}{\mathcal{J}}
\newcommand{\calK}{\mathcal{K}}
\newcommand{\calO}{\mathcal{O}}
\newcommand{\Fig}[2]{\includegraphics[width=#1]{#2}}
\renewcommand{\Re}{\mathrm{Re}}
\renewcommand{\Im}{\mathrm{Im}}

\usepackage{color}
\usepackage[normalem]{ulem}

\renewcommand\sout{\bgroup \color{red} \ULdepth=-.5ex \ULset}

\begin{document}

\title{An improvement in complex Langevin dynamics from a view point of Lefschetz thimbles}

\author{Shoichiro Tsutsui}
\email[]{tsutsui@ruby.scphys.kyoto-u.ac.jp}
\affiliation{Department of Physics, Faculty of Science, Kyoto University,
	Kyoto 606-8502, Japan}
\author{Takahiro M. Doi}
\email[]{doi@ruby.scphys.kyoto-u.ac.jp}
\affiliation{Department of Physics, Faculty of Science, Kyoto University,
	Kyoto 606-8502, Japan}

\date{\today}

\begin{abstract}
We develop a way of improving complex Langevin dynamics motivated by the Lefschetz-thimble decomposition of integrals.
In our method, arbitrary observables of an original model with multiple Lefschetz thimbles are computed by a modified model with a single thimble.
We apply our modification method to a one dimensional integral 
in which the naive implementation of the complex Langevin dynamics fails to reproduce the exact results due to the severe sign problem.
We show that the toy model can be modified so that the new model consists of a single Lefschetz thimble.
We find that correct results can be obtained by the improved complex Langevin dynamics.
\end{abstract}

\maketitle

\section{Introduction}
The quantum Monte Carlo simulation is a sophisticated way to reveal nonperturbative physics for theories with real actions.
However, it does not work well for theories with complex actions due to the breakdown of the interpretation of the Boltzmann factor $e^{-S}$ as a probability distribution.
Complex actions are found in many fields of physics; frustrated spin systems, the Hubbard model away from half filling, QCD at finite chemical potential and all real time problems.
In these cases, the quantum Monte Carlo simulation suffers from the sign problem which prevents us from measuring physical quantities with appropriate precision~\cite{Loh:1990,vonderLinden199253,Muroya:2003qs,deForcrand:2010ys}.
To establish first-principle approaches for complex actions is still an outstanding problem in physics.

To circumvent the sign problem, the stochastic quantization is explored to apply theories with complex actions.
The framework of the ordinary stochastic quantization is established for theories with real actions, 
where a quantum average is computed by solving a Langevin equation~\cite{Parisi:1980ys,Damgaard:1987rr,Namiki:1992}.
At least formally, the stochastic quantization seems to be extended to cases with complex actions~\cite{Parisi:1984cs,Klauder:1983,Klauder:1983sp,Klauder:1985,Ambjorn:1985iw}.
However, the complex Langevin dynamics sometimes gives incorrect results~\cite{Ambjorn:1986fz,Flower:1986hv}.
Recently, a necessary and sufficient condition for correctness of the complex Langevin dynamics is proposed and a formal justification of this method is established~\cite{Aarts:2009uq,Aarts:2011ax}.
Nevertheless, it is still difficult to expect a priori when the complex Langevin dynamics gives correct answers,
although practical methods are proposed to improve it~\cite{Aarts:2009dg,Seiler:2012wz,Mollgaard:2014mga}.
Moreover, there are subtleties when an action involves logarithmic terms.
For instance, such a logarithmic term comes from a fermionic determinant in QCD at finite chemical potential.
The logarithmic term leads to a singular behavior of a drift term and thus the complex Langevin dynamics can fail to give correct results~\cite{Mollgaard:2013qra,Greensite:2014cxa,Nishimura:2015pba}.

Another possible way to overcome the sign problem is the Lefschetz-thimble decomposition which is a generalization of the steepest descent method.
In that framework, an integration path is deformed to a set of curved manifolds in a complex space which are called Lefschetz thimbles.
This method is extended to higher dimensions~\cite{Pham:1983} and directly applied to the quantum field theories~\cite{Witten:2010cx}.
The numerical implementation of this method is also discussed extensively~\cite{Cristoforetti:2012su,Cristoforetti:2013wha,Fujii:2013sra,Mukherjee:2014hsa,Tanizaki:2014tua,DiRenzo:2015foa,Fukushima:2015qza,Tanizaki:2015rda,Fujii:2015bua,Fujii:2015vha,Alexandru:2015xva}.
The method of Lefschetz thimbles is well established, but in principle, non-holomorphic actions are beyond the scope of this method.
Nevertheless, for known cases, Lefschetz thimbles are well defined for actions with logarithmic terms~\cite{Kanazawa:2014qma}.
The relation between complex Langevin dynamics and Lefschetz thimbles is first discussed in~\cite{Aarts:2013fpa,Aarts:2014nxa}.
It is suggested that in some cases configurations sampled by the complex Langevin process can be interpreted as the importance sampling on Lefschetz thimbles.

In this paper, we consider the cases where the configurations generated by the complex Langevin process are sampled on Lefschetz thimbles.
In such a case, we point out that multi-thimble structure causes the wrong convergence of complex Langevin simulations.
We develop a way of improving complex Langevin dynamics by modifying a structure of Lefschetz thimbles and demonstrate how our approach works by applying it to the cosine model which has a severe sign problem due to the logarithmic term in its action analogous to the chiral random matrix theory and QCD at finite chemical potential.

\section{Complex Langevin dynamics and Lefschetz thimbles}
In this section, we briefly introduce the two frameworks, complex Langevin dynamics and the Lefschetz-thimble decomposition.

In the ordinary stochastic quantization, it is assumed that dynamical variables depend on a fictitious time $t$ and satisfy the Langevin equation.
Once the system relaxes to thermal equilibrium, the noise average is interpreted as the quantum average.
This framework is formally generalized to theories with complex actions.
Supposing that the action of an original theory is given by $S(x)$, where $x\in\mathbb{R}$ and $S(x)\in\mathbb{C}$,
the stochastic quantization is applied by replacing $x$ with $z\in\mathbb{C}$.  
Then, the Langevin equation with a complex action $S(z)$ is given by
\begin{align}
	\frac{\del z}{\del t} = - \frac{\del S(z)}{\del z} + \eta,
	\label{CLE}
\end{align}
where $\eta$ is a real Gaussian noise satisfying $\braket{\eta(t)\eta(t^\prime)}=2\delta(t-t^\prime)$.
$\braket{\dots}$ indicates the noise average here.

The other useful framework is the Lefschetz-thimble decomposition.
Lefschetz thimbles $\calJ_\sigma$ are defined as the paths of steepest descent starting from saddle points of the action $z_\sigma$, where $\sigma$ is a label of each saddle point. 
The steepest descent path is determined by the following flow equation:
\begin{align}
	\frac{\del z}{\del t} = - \overline{ \lp \frac{\del S(z)}{\del z} \rp} .
	\label{thimble}
\end{align}
Note that the saddle point $z_\sigma$ is nothing but a fixed point of the flow equation.
The steepest ascent path $\calK_\sigma$ starting from the same point as $\calJ_\sigma$ is also determined by a similar flow equation whose sign of the gradient term is opposite to Eq.~\eqref{thimble}.
By utilizing the set of Lefschetz thimbles, the partition function is decomposed as follows:
\begin{align}
	Z = \int dx e^{-S(x)} = \sum_\sigma n_\sigma e^{-i \Im S(z_\sigma)} \int_{\calJ_\sigma} dz e^{-\Re S(z)} .
	\label{decomposition}
\end{align}
The global sign factor $e^{-i\Im S(z_\sigma)}$ is factorized outside the integral since the imaginary part of the action is constant along each Lefschetz thimble.
Coefficient $n_\sigma$ is the number of intersections of the original integration path and the steepest ascent path $\calK_\sigma$.
We refer to a Lefschetz thimble which has non-zero intersection number as a relevant thimble.

There are various arguments about relations between the complex Langevin dynamics and the Lefschetz-thimble decomposition.
Some cases are known that the complex Langevin simulation is regarded as an importance sampling on Lefschetz thimbles while the opposite cases are also known. In this paper, we consider the cases where the configurations generated by the complex Langevin process are sampled on Lefschetz thimbles in order to identify a cause of wrong convergence problem of the complex Langevin simulation from a viewpoint of the Lefschetz thimbles.
It may be naively expected that the complex Langevin dynamics gives correct answers when the importance sampling on Lefschetz thimbles is achieved.
However, this is neither necessary nor sufficient for the correctness of the complex Langevin dynamics~\cite{Aarts:2013fpa,Aarts:2013uza}.
In fact, as we see below, the complex Langevin simulation does not give correct answer if there are more than two Lefschetz thimbles.

\section{Cosine model}
Let us consider a toy model whose partition function is given by the one dimensional integral,
\begin{align}
Z = \int_{-\pi}^{\pi} dx \cos x e^{\beta \cos x} .
	\label{cosmodel}
\end{align}
This cosine model is obtained by the factorization of the two dimensional U(1) model~\cite{Ambjorn:1986fz}.
The action of this model reads $S = -\beta \cos x - \log(\cos x)$
and it is complex due to the logarithmic term when $\pi/2 < |x| < \pi$.
We assume that the parameter $\beta$ is real and positive.
The expectation value of $\cos nx$ can be expressed analytically:
\begin{align}
\braket{\cos nx}_Z 
&= 
\frac{\int_{-\pi}^\pi dx \cos nx \cos x e^{\beta \cos x}}{\int_{-\pi}^\pi dx  \cos x e^{\beta \cos x}} \notag \\
&=
\frac{I_{n+1}(\beta)}{I_1(\beta)} + \frac{n}{\beta} \frac{I_{n}(\beta)}{I_1(\beta)} ,
\label{exact}
\end{align}
where $I_n(\beta)$ is the modified Bessel function of the first kind.
Note that in $n=1$ case, the expectation value $\braket{\cos x}_Z$ diverges as $\sim 1/\beta$ for $\beta \ll 1$.

Concerning the sign problem, it is instructive to consider this toy model.
In fact, the reweighting technique, a way to tame the sign problem, does not give correct answers for this model since the average phase factor $\Braket{\cos(x)/|\cos(x)|} $ becomes much smaller than 1 as $\beta$ goes to zero \cite{deForcrand:2010ys,Mollgaard:2013qra}.
These behaviors are reminiscent of the sign problem in QCD at finite chemical potential.
Moreover, as discussed in Ref. \cite{Ambjorn:1986fz}, the complex Langevin dynamics also fails to reproduce the exact result Eq.~\eqref{exact} for the cosine model. (This behavior is also found in Fig.~\ref{Fig:expectation}.)
In the following, we discuss how the sign problem appears in the cosine model for $\beta<1$.

\begin{figure}[th]
	\centering
	\Fig{8.5cm}{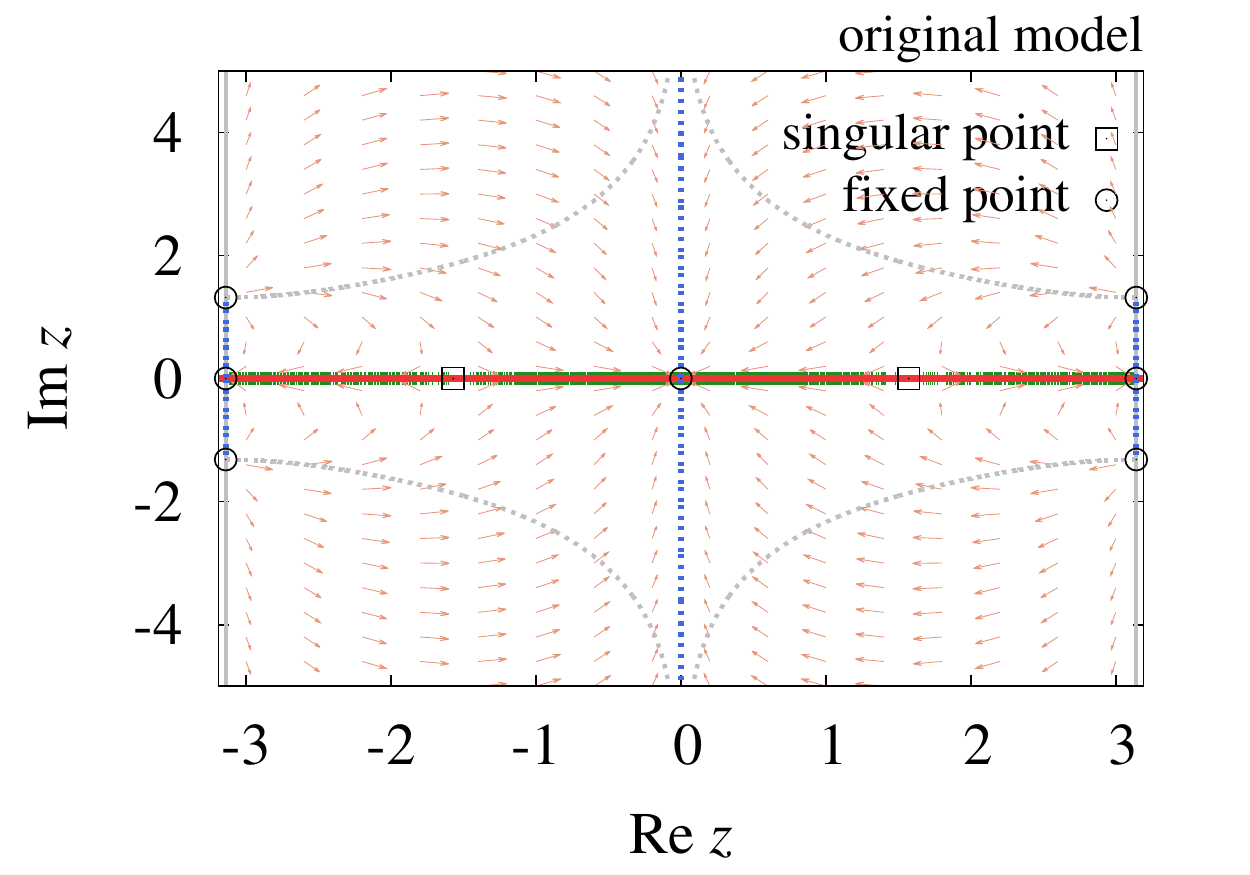}
	\Fig{8.5cm}{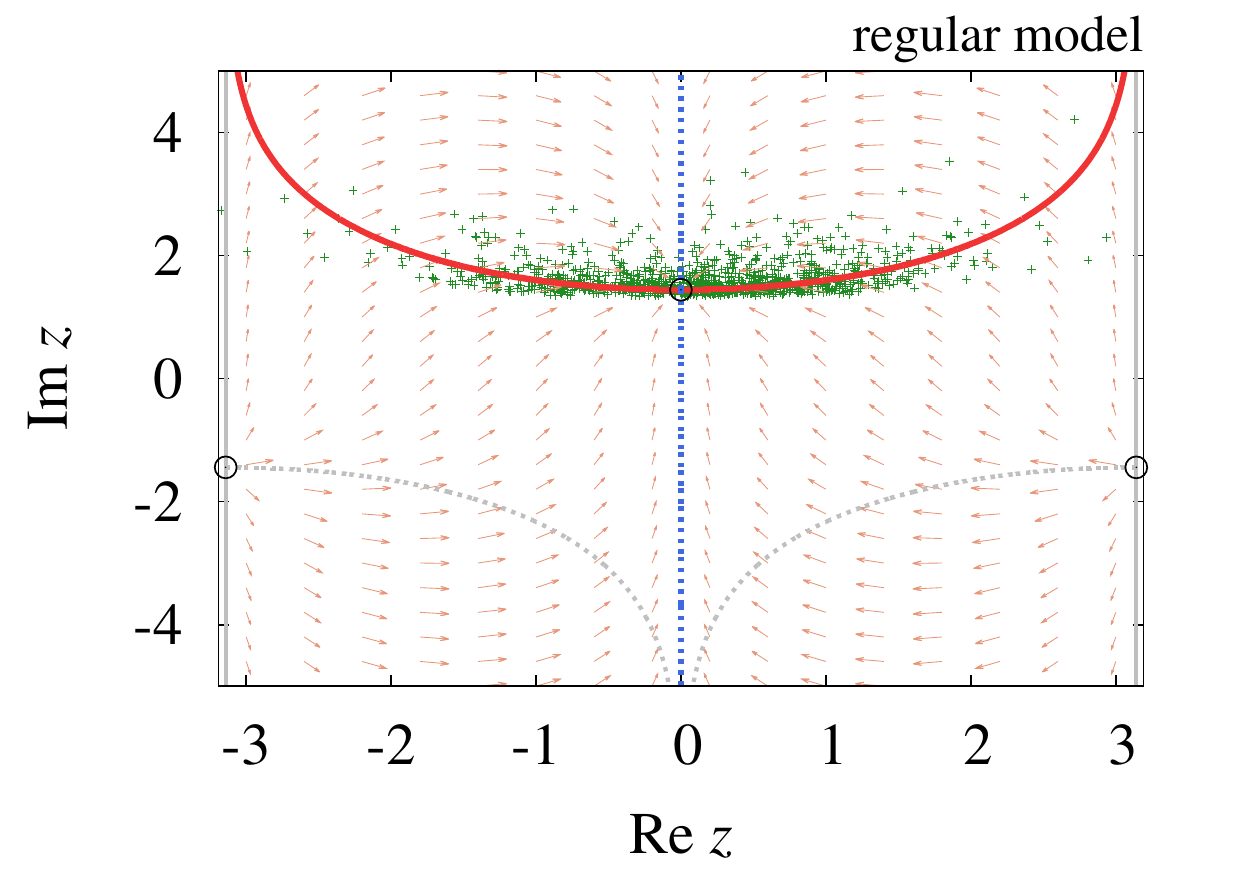}
	\caption{
		Lefschetz thimbles and sampled configurations of the original (upper panel) and the regular (lower panel) cosine models for $\beta=0.5$.
		Solid (dotted) lines are steepest descent (ascent) paths. 
		Arrows denote the drift term of Eq.~\eqref{CLE}.
	}
	\label{Fig:orgthimbles}
\end{figure}
In Fig.~\ref{Fig:orgthimbles}, the Lefschetz thimbles and the corresponding steepest ascent paths are denoted by solid and dotted lines for $\beta=0.5$.
The scattered data are configurations sampled by the complex Langevin process.
We also illustrate the drift term of the complex Langevin equation by arrows.
One finds that the configurations seem to be well distributed around the relevant thimbles.
This result suggests that the complex Langevin dynamics may not work even when an importance sampling on Lefschetz thimbles is achieved. To find the origin of the failure we focus on the structure of Lefschetz thimbles.

\begin{figure}[bth]
	\centering
	\Fig{8.5cm}{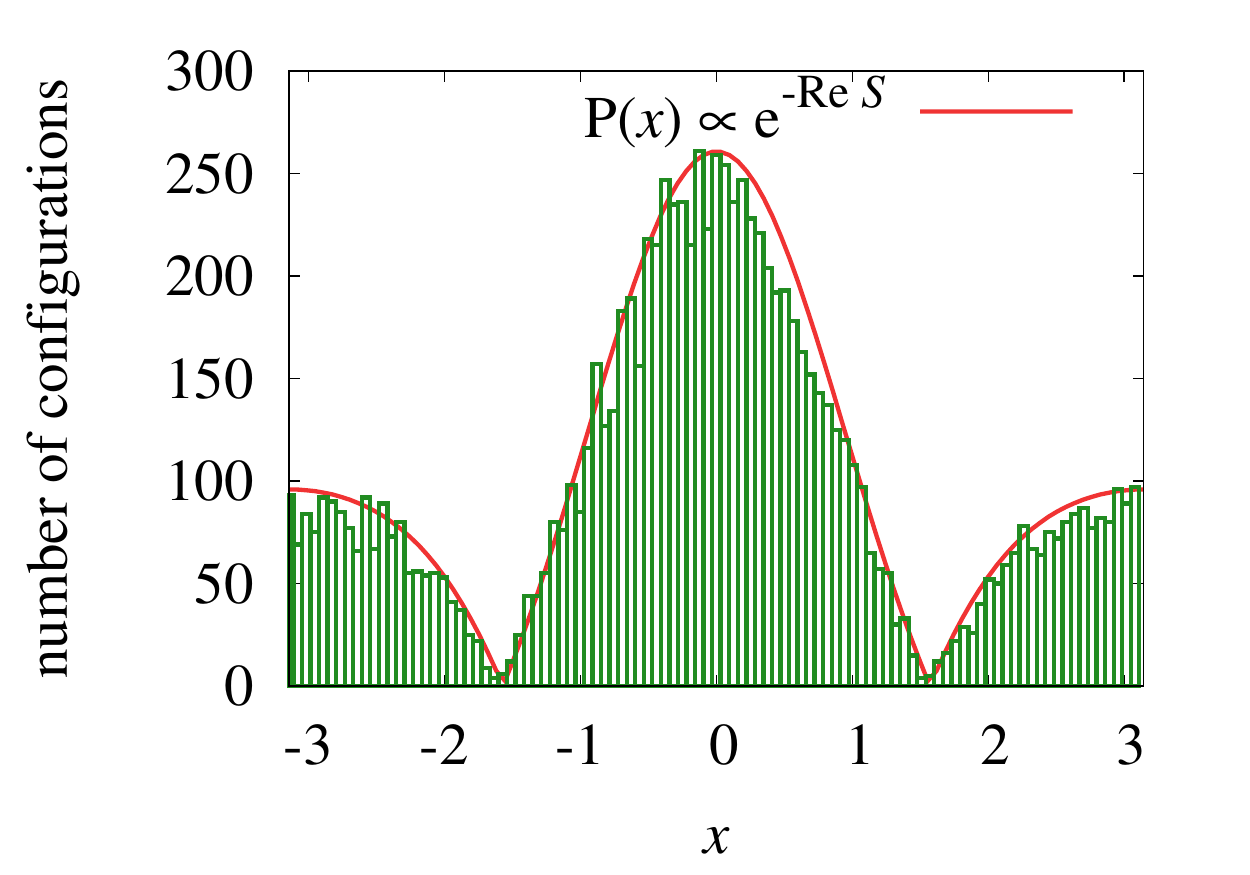}
	\caption{
		The distribution of the original cosine model obtained by the complex Langevin process for $\beta=0.5$.
		The solid line indicates the distribution function $P(x) \propto e^{-\Re S(x)}$.
	}
	\label{Fig:hist_org}
\end{figure}

In the cosine model, there exist four fixed points $z_f = 0$, $\pi$ and $\pi \pm i \cosh^{-1}(1/\beta)$, while only the two thimbles starting from $z_f = 0$ and $\pi$ contribute to the partition function Eq~\eqref{cosmodel} when $\beta < 1$.
The values of the action on these fixed points are $S(0)=-\beta$ and $S(\pi)=\beta-i\pi$.
Decreasing $\beta$, the real parts of these values come close, and then, the contribution from each thimble should cancel out.
This situation is referred to as the global sign problem in the context of the method of Lefschetz thimbles.
Despite the cancellation of multi-thimble contribution should be crucial to reproduce correct results \cite{Tanizaki:2015rda}, complex Langevin processes do not take into account the phase factor $e^{-i\Im S(z_f)}$ automatically \cite{Mollgaard:2013qra,Hayata:2015lzj}.
Figure~\ref{Fig:hist_org} shows the histogram of configurations sampled by the complex Langevin process for $\beta=0.5$.
Each configuration holds $\Im z\simeq0$, which means that all sampled points are contained in the Lefschetz thimbles.
The histogram has two peaks around $z=0$ and $\pi$, and its shape is characterized by the distribution $P(x) \propto e^{-\Re S(x)}$ without any information of the imaginary part of the action.
This can spoil the validity of the complex Langevin processes.

The interesting feature of the cosine model is that we can improve the complex Langevin process so that it gives correct answers \cite{Ambjorn:1986fz}.
Recalling that the partition function Eq.~\eqref{cosmodel} is invariant by adding $0=i\int_{-\pi}^{\pi} dx \sin x e^{\beta \cos x}$,
one finds
\begin{align}
	Z = \int_{-\pi}^{\pi} dx e^{\beta \cos x + i x},
	\label{regcosmodel}
\end{align}
which is equivalent to Eq.~\eqref{cosmodel}.
The action obtained from the new expression reads $S_\text{reg} = -\beta \cos x - i x$ which does not involve logarithmic singularities.
We refer to this model as the regular cosine model. 
The lower panel of Fig.~\ref{Fig:orgthimbles} shows the Lefschetz thimbles of the regular cosine model for $\beta = 0.5$.
We find that the original integration path is deformed to the single Lefschetz thimble and the configurations sampled by the complex Langevin process are well distributed around it.
In contrast to the original cosine model, the phase factor $e^{-i\Im S(z_f)}$ does not matter for the regular cosine model since it consists of only one Lefschetz thimble.
Indeed, the complex Langevin process reproduces analytic results as expected.

These two models may give insights to improve the complex Langevin processes.
To be specific, it is suggested that the complex Langevin processes can be cured by modifying the partition function so that the process is interpreted as an importance sampling on a \textit{single} Lefschetz thimble.
On the other hand, it seems hopeless to find such modifications for more complicated models.
In particular, in the derivation of the regular cosine model, the added integral $0=i\int_{-\pi}^{\pi} dx \sin x e^{\beta \cos x}$ is apparently fine-tuned so that the partition function consists of one Lefschetz thimble.

In the following, we generalize the procedure to construct a theory with single-thimble structure and give an example to test our idea.

\section{Modification of Lefschetz thimbles}
We propose a method to modify an original theory to the theory with a single Lefschetz thimble.
Let us consider a partition function which has the following form;
\begin{align}
	Z \equiv Z[f(x)] = \int_{D_0} dx f(x) e^{-S_\text{q}(x)} ,
	\label{original}
\end{align}
where $D_0$ is a domain of integration.
We assume that $D_0$ have a finite width. 
The total action is given by $S = S_\text{q} - \log f$ and the quenched part $S_\text{q}$ is assumed to be real.
We define the modification of $Z$ by
\begin{align}
	Z[f+g] = \int_{D_0} dx (f(x) + g(x)) e^{-S_\text{q}(x)}.
\label{modified}
\end{align}
$g$ is an arbitrary holomorphic function.
We also define the quenched partition function $Z_\text{q} = Z[f=1]$.
One can verify the following identity for an arbitrary observable $\calO(x)$;
\begin{align}
	\braket{\calO}_{Z}
	=
	\braket{\calO}_{Z[f+g]} + \lp \braket{\calO}_{Z[f+g]} - \braket{\calO}_{Z[g]} \rp \frac{ \braket{g}_{Z_\text{q}} }{ \braket{f}_{Z_\text{q}} } .
	\label{general connection formula}
\end{align}
$\braket{\calO}_{Z}$ denotes the expectation of $\calO$ for a partition function $Z$;
\begin{align}
	\braket{\calO}_{Z} &= \frac{\int dx \calO(x) f(x) e^{-S_\text{q}(x)}}{\int dx f(x) e^{-S_\text{q}(x)}}.
\end{align}
Expectation values for the modified and quenched partition functions are defined in the same way. 
Thus, any expectation values of the original theory can be computed from the modified and quenched theories.
In particular, $\braket{f}_{Z_\text{q}}$ and $\braket{g}_{Z_\text{q}}$ are calculable on the basis of the ordinary Langevin method or the Monte Carlo simulation because the $S_\text{q}$ is supposed to be real. 
Remarkably, Eq.~\eqref{general connection formula} holds in a higher dimensional case as well as the present one-dimensional integral.

It is worth to mention a case where $\braket{g}_{Z_\text{q}}=0$. In such a case, $\braket{f}_{Z_\text{q}}$ never contribute to the $\braket{\calO}_{Z}$ even when the sign problem is severe, or $\braket{f}_{Z_\text{q}}$ is hard to compute by the Monte Carlo simulation. The regular cosine model discussed around Eq.~\eqref{regcosmodel} is an example of such a modification.  We also emphasize that our modification method does not need to eliminate the logarithmic terms from the original model like the case of the regular cosine model. 

For the sake of later discussion,
we give a useful identity which is rather simpler than Eq.~\eqref{general connection formula}.
This identity is obtained by the replacement $g(x) \to ig(x)$,
where the new $g(x)$ is holomorphic, and takes only real values on the real axis.
For such a restricted class of $g$ and real function $\calO$, we find 
\begin{align}
	\braket{\calO}_{Z}
	=
	\Re\braket{\calO}_{\tilZ} - \frac{\braket{g}_{Z_\text{q}}}{\braket{f}_{Z_\text{q}}	} \Im\braket{\calO}_{\tilZ} .
	\label{connection formula}
\end{align}
Here $\tilZ$ denotes $Z[f+ig]$.
As will see below, 
this simpler identity is sufficient to get a single-thimble theory in the case of the cosine model.

The next step to see is how the function $g(x)$ modifies the Lefschetz thimbles of the original theory.
The shape of Lefschetz thimbles is characterized by fixed and singular points which are the starting and ending points of the steepest descent path.
The fixed point $z_f$ and the singular point $z_s$ of the modified theory Eq.~\eqref{modified} are obtained by the following equations:
\begin{align}
& f(z_s) + i g(z_s) = 0, \label{singlarity}\\
& \lp S_\text{q}^\prime(z) - \frac{f^\prime(z) + i g^\prime(z)}{f(z) + i g(z)} \rp\Bigg|_{z=z_f} = 0,\label{fixed}
\end{align}
where primes denote the derivative with respect to $z$.
In general, it is difficult to find all the solutions of Eqs.~\eqref{singlarity} and~\eqref{fixed} for an arbitrary $g(x).$ To avoid the difficulty, we introduce a real and positive parameter $\tau$ and replace $g(z)$ by $\tau g(z)$. When $\tau=0$, one reproduces the original partition function Eq.~\eqref{original}. The location of the fixed and singular points are described by the following evolution equations which are much easier to solve Eqs.~\eqref{singlarity} and~\eqref{fixed}:
\begin{align}
&\frac{dz_s}{d\tau}
=
\frac{-ig(z_s)}{f^\prime(z_s)+i\tau g^\prime(z_s)},
\label{evol_sing} \\
%
&\frac{dz_f}{d\tau} \frac{d}{dz} \lp S_\text{q}^{\prime}(z) - \frac{f^{\prime}(z)+i\tau g^{\prime}(z)}{f(z)+i\tau g(z)} \rp \Bigg|_{z=z_f} = 0. 
\label{evol_fix}		 
\end{align}

Equation~\eqref{evol_sing} is also useful to determine a form of $g(z)$ which leads singe-thimble structure.
If there are singular points in the domain of the complex action $S(z)$ and they are the endpoints of relevant thimbles, there are more than two thimbles.
The necessary condition for single-thimble structure to achieve is that the singular points are located outside or on the boundary of the domain.
Since a zero of $g(z)$ is a fixed point of Eq.~\eqref{evol_sing}, the zero of $g(z)$ is a singular point of the modified model for a sufficiently large $\tau$.
Therefore, $g(z)$ should have zeros outside or on the boundary of the domain.

While this is a criterion for choosing $g(z)$, it does not always guarantee that the modified model consists of a single Lefschetz thimble.
Specifically, zeros of $g(z)$ are not necessarily the endpoints of relevant thimbles.
It should be confirmed for each $g(z)$ whether single-thimble structure is achieved or not in the modified model.

\section{Application to the Cosine model}
In this section, we apply the modification method to the cosine model, namely,
$f(x) = \cos x$, $S_\text{q}=-\beta \cos x$ and $D_0 = [-\pi,\pi]$ in Eq.~\eqref{original}.
The domain of the complex action $S(z)$ is given by $|{\rm Re}z|\leq \pi$.
As we can see in Fig.~\ref{Fig:orgthimbles}, the cosine model has the multi-thimble structure for $\beta = 0.5$.
The multi-thimble structure is caused by the two singular points, $z = \pm \pi/2$ because they are the endpoints of the relevant thimbles.

\begin{figure}[bth]
	\centering
	\Fig{8.5cm}{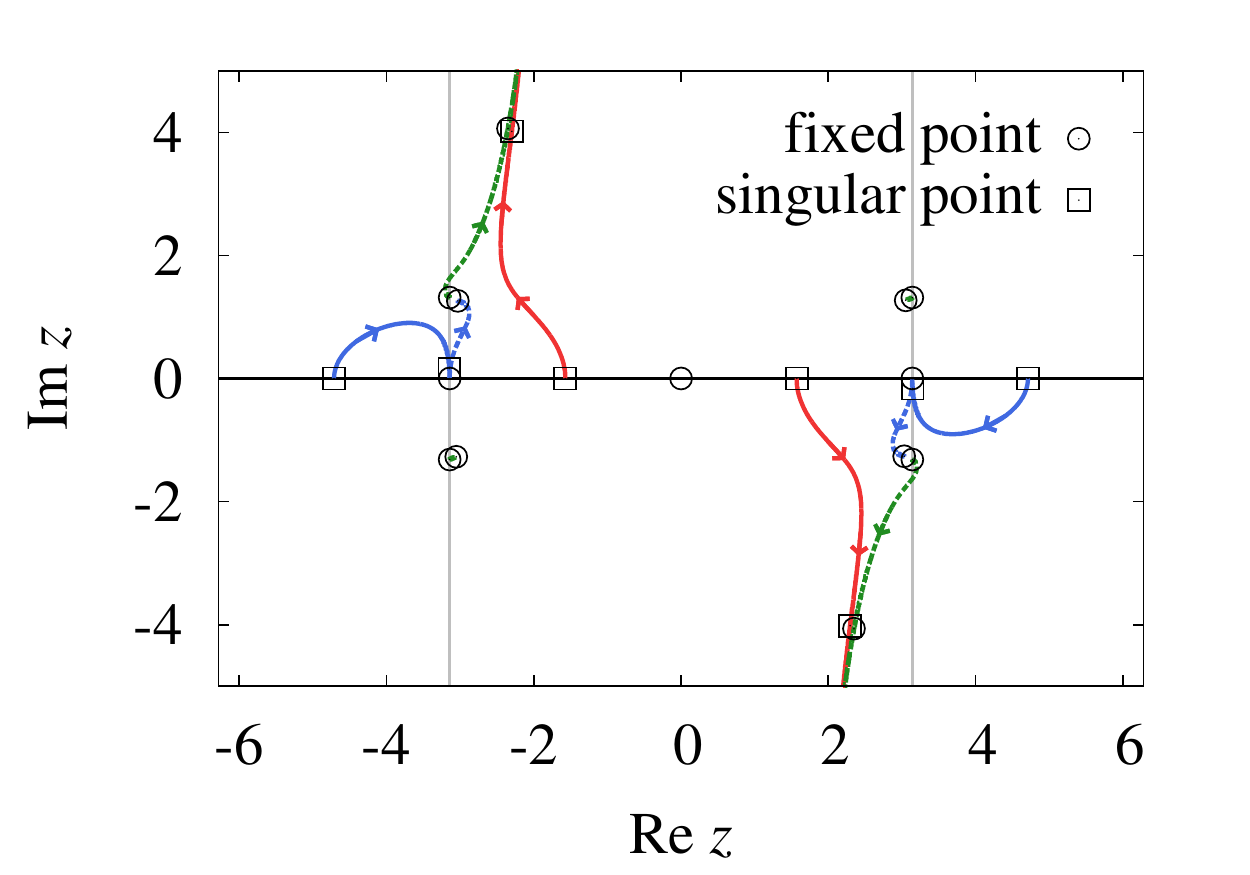}
	\Fig{8.5cm}{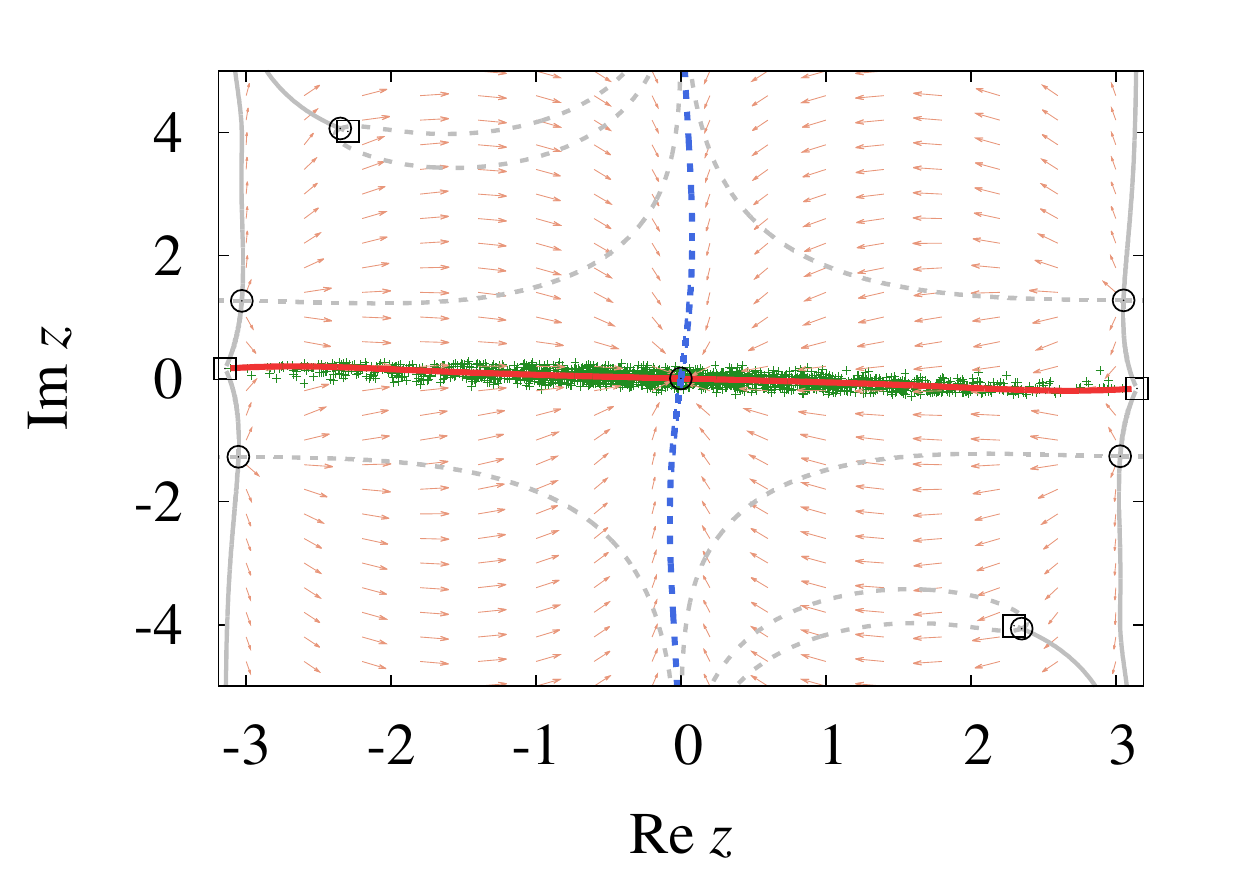}
	\caption{
		(top) The evolution of fixed and singular points for $\beta=0.5$.
		Each point is plotted for $\tau = 0$ and 1. 
		The function $g(z)$ is taken as $ g(z)=(z-\pi)(z+\pi)$.
		(bottom) The Lefschetz thimbles at $\tau=1$.
		The sampled configurations of the complex Langevin process are plotted for the periodic boundary condition.
	}
	\label{Fig:evol}
\end{figure}

In order to achieve single-thimble structure, it is necessary that the relevant thimble starting from the origin has the endpoints outside or on the boundary of the domain $|{\rm Re}z|\leq \pi$.
When we impose $g(\pm\pi)=0$, the minimal choice of $g(z)$ is given by $g(z)=(z-\pi)(z+\pi)$.
In Fig.~\ref{Fig:evol}, we show the evolution of the singular and fixed points from $\tau=0$ to 1 for $\beta = 0.5$.
The endpoints of the thimble of the original model at $z = \pm \pi/2$ leave from the real axis while the singular points at $z = \pm 3\pi/2$ move towards $z=\pm \pi$.
We also show the Lefschetz thimbles at $\tau=1$ in Fig.~\ref{Fig:evol}.
Only the thimble starting from the origin is relevant, and it ends at $z=\pm \pi$ as we expected.
Of course, $g(z)$ is not unique and other choices of $g(z)$ can lead single-thimble structure, for example, $g(z)= (z-2\pi)(z+2\pi)$. 
We have also confirmed that the integration contour of the modified model consists of one thimble for that case.
In the following, we employ $ g(z)=(z-\pi)(z+\pi)$ and $\tau=1$ to perform the complex Langevin simulation.

The complex Langevin equation of the modified model is solved by the usual Euler scheme.
Because $g(z) = (z-\pi)(z+\pi)$ breaks the periodicity of the original model,
we use the periodic and box boundary conditions so that configurations are sampled in the domain $|\Re z| \leq \pi$.
In the box boundary condition, the drift term is modified to add the repulsive barrier at the boundary of the area. The drift term including the repulsive force is given by $\del S/ \del z - \delta(\Re z - \pi) + \delta(\Re z + \pi)$.
The time slice is $\Delta t = 10^{-4}$ and the total Langevin step is $N = 10^9$.
We use an adaptive time step~\cite{Flower:1986hv} to stabilize the numerical simulation.
$\braket{f}_{Z_\text{q}}$ and $\braket{g}_{Z_\text{q}}$ are measured by the Monte Carlo integration.
We measure the expectation value of $\cos nx$ by utilizing Eq.~\eqref{connection formula}.

\begin{figure}[bth]
	\centering
	\Fig{8.5cm}{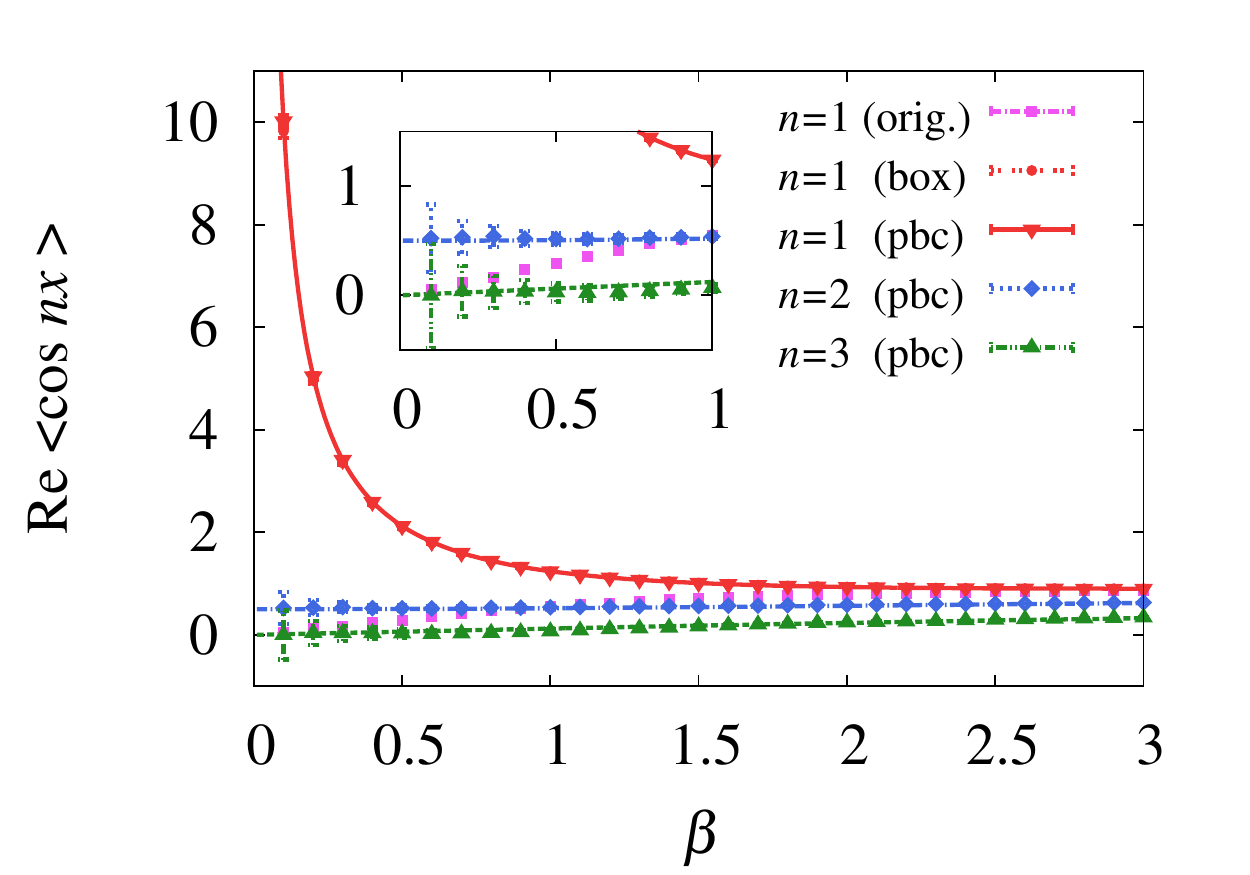}
	\caption{
		The expectation values of $\Re \braket{\cos nx}_Z$ computed from original (denoted by square points) and modified cosine models.
		For the 	modified model, the results are shown for box and periodic boundary conditions (pbc).
		The analytic results Eq.~\eqref{exact} are plotted by solid lines.
	}
	\label{Fig:expectation}
\end{figure}

Figure~\ref{Fig:expectation} shows the value of $\Re \braket{\cos nz}_Z$ as a function of $\beta$.
The expectation values measured by the modified model well agree with the analytic results Eq.~\eqref{exact} within the statistical error.
Specifically, the $1/\beta$ behavior of $\braket{\cos x}_Z$ in the small $\beta$ region is clearly reproduced while
the complex Langevin dynamics of the original cosine model fails.
We confirmed that the choice of the boundary conditions does not affect the result.

\begin{figure}[bth]
	\centering
	\Fig{8.5cm}{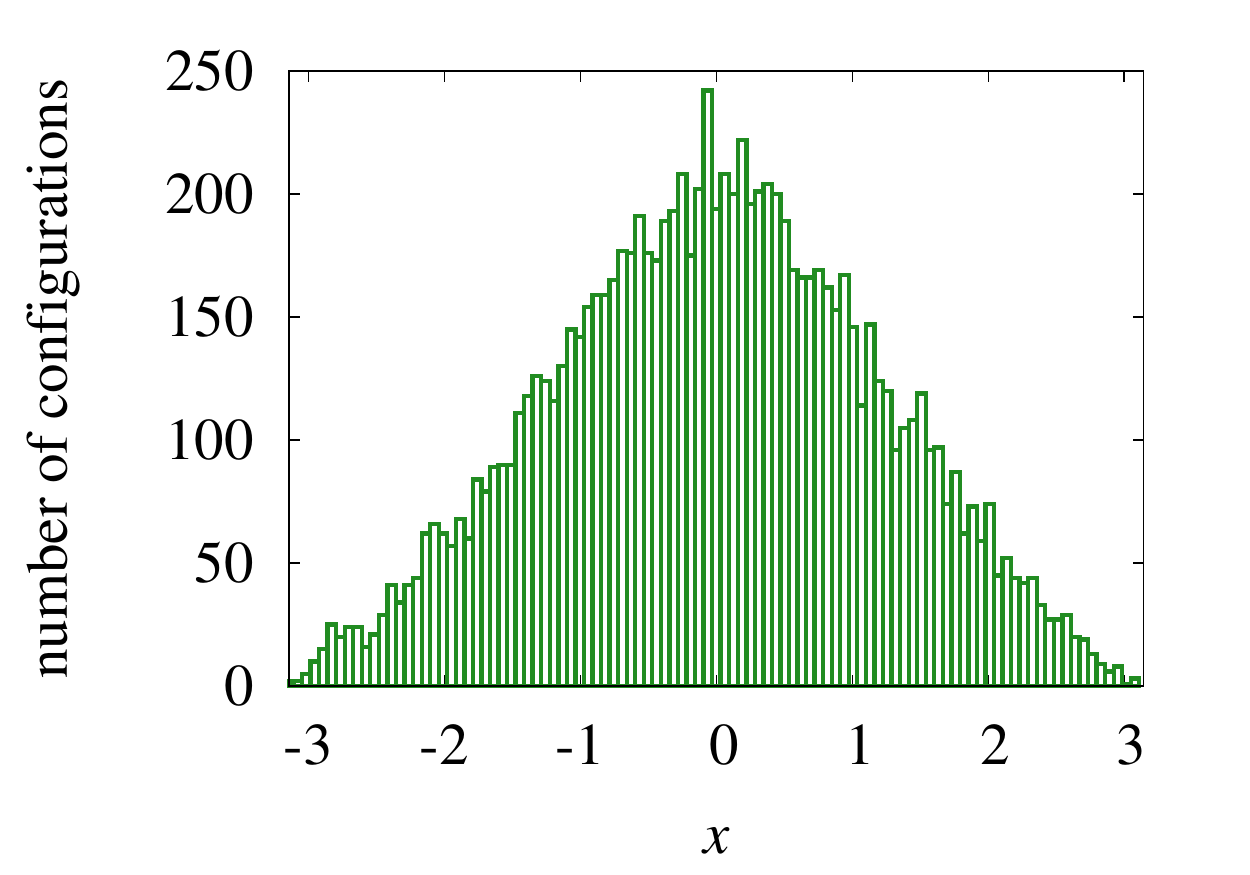}
	\caption{
		The distribution obtained by the complex Langevin process.
	}
	\label{Fig:thimbles}
\end{figure}

Finally, we discuss the distribution obtained by the complex Langevin process for the modified model.
Figure~\ref{Fig:thimbles} shows the histogram of configurations sampled by the complex Langevin process with the periodic boundary condition for $\beta=0.5$.
Since the origin is the attractive point under the Langevin dynamics,
the configurations are manifestly distributed around the Lefschetz thimble.
The choice of the boundary condition does not change the shape of the distribution qualitatively.

Concerning the validity of the complex Langevin dynamics, we should mention the role of singular points at $z=\pm\pi$.
In Refs.~\cite{Mollgaard:2013qra,Greensite:2014cxa,Nishimura:2015pba},
it was argued that the failure of the complex Langevin dynamics may be attributed to the presence of singular points.
Nevertheless, we find that the expectation values are correctly measured while the distribution seems to overlap the singular points.

\section{Conclusions}
We have developed a way to improve the complex Langevin dynamics by modifying the structure of Lefschetz thimbles.
The desirable modification can be found by manipulating the locations of fixed and singular points.
We have applied our method to solve the cosine model in which a severe sign problem arises to test its validity.
As a result, it has been found that the expectation value can be correctly measured by the complex Langevin process when the modified partition function consists of a single Lefschetz thimble and the process achieves the importance sampling on it.

One of the most interesting application of our method is dealing with phase transitions.
In the vicinity of a transition line, both perturbative and nonperturbative fixed points can contribute to the partition function, and therefore, it consists of multi-thimbles. 
Even in the situation, one can manipulate the structure of the Lefschetz thimbles such that only one thimble is relevant.
It is also interesting to find a modification function $g(x)$ which holds $\braket{g(x)}_{Z_\text{q}}=0$ and achieves an importance sampling on a single Lefschetz thimble in more complicated models.
In such a case, the expectation value measured in the original model $\braket{\calO(x)}$ can be obtained without calculating $\braket{f(x)}_{Z_\text{q}}$ appearing in Eqs.~\eqref{general connection formula} and \eqref{connection formula} which is hard to compute reliably in systems with a severe sign problem.

Other important issues are higher dimensional problems where the structure of Lefschetz thimbles is more complicated.
To develop more systematic ways to find an appropriate modification is work in progress.

\section*{Acknowledgments}
We thank T. Ichihara, K. Kashiwa, A. Ohnishi, S. Shimasaki, Y. Tanizaki and H. Tsukiji for fruitful discussions.
S.T. is supported by the Grant-in-Aid for JSPS fellows (No.26-3462).
T.M.D. is supported by the Grant-in-Aid for JSPS fellows (No.15J02108).

\bibliographystyle{h-physrev5}
\bibliography{CLEref.bib}

\begin{thebibliography}{10}

\bibitem{Loh:1990}
E.Y.~Loh, J.E.~Gubernatis, R.T.~Scalettar, S.R.~White, D.J.~Scalapino, R.L.~Sugar,
\newblock Phys. Rev. B {\bf 41}, 9301 (1990).

\bibitem{vonderLinden199253}
W.~von~der Linden,
\newblock Physics Reports {\bf 220}, 53  (1992).

\bibitem{Muroya:2003qs}
S.~Muroya, A.~Nakamura, C.~Nonaka, and T.~Takaishi,
\newblock Prog. Theor. Phys. {\bf 110}, 615 (2003), arXiv:hep-lat/0306031.

\bibitem{deForcrand:2010ys}
P.~de~Forcrand,
\newblock PoS {\bf LAT2009}, 010 (2009), arXiv:1005.0539.

\bibitem{Parisi:1980ys}
G.~Parisi and Y.-S. Wu,
\newblock Sci. Sin. {\bf 24}, 483 (1981).

\bibitem{Damgaard:1987rr}
P.~H. Damgaard and H.~Huffel,
\newblock Phys. Rept. {\bf 152}, 227 (1987).

\bibitem{Namiki:1992}
M.~Namiki,
\newblock Lect. Notes Phys. M {\bf 9}, p. 1 (1992).

\bibitem{Parisi:1984cs}
G.~Parisi,
\newblock Phys. Lett. {\bf B131}, 393 (1983).

\bibitem{Klauder:1983}
J.~Klauder,
\newblock Stochastic quantization,
\newblock in {\em Recent Developments in High-Energy Physics}, edited by
  H.~Mitter and C.~Lang, , Acta Physica Austriaca Vol. 25/1983, pp. 251--281,
  Springer Vienna, 1983.

\bibitem{Klauder:1983sp}
J.~R. Klauder,
\newblock Phys. Rev. {\bf A29}, 2036 (1984).

\bibitem{Klauder:1985}
J.~R. Klauder,
\newblock J. Stat. Phys. {\bf 39}, 53 (1985).

\bibitem{Ambjorn:1985iw}
J.~Ambjorn and S.~K. Yang,
\newblock Phys. Lett. {\bf B165}, 140 (1985).

\bibitem{Ambjorn:1986fz}
J.~Ambjorn, M.~Flensburg, and C.~Peterson,
\newblock Nucl.Phys. {\bf B275}, 375 (1986).

\bibitem{Flower:1986hv}
J.~Flower, S.~W. Otto, and S.~Callahan,
\newblock Phys.Rev. {\bf D34}, 598 (1986).

\bibitem{Aarts:2009uq}
G.~Aarts, E.~Seiler, and I.-O. Stamatescu,
\newblock Phys.Rev. {\bf D81}, 054508 (2010), arXiv:0912.3360.

\bibitem{Aarts:2011ax}
G.~Aarts, F.~A. James, E.~Seiler, and I.-O. Stamatescu,
\newblock Eur.Phys.J. {\bf C71}, 1756 (2011), arXiv:1101.3270.

\bibitem{Aarts:2009dg}
G.~Aarts, F.~A. James, E.~Seiler, and I.-O. Stamatescu,
\newblock Phys.Lett. {\bf B687}, 154 (2010), arXiv:0912.0617.

\bibitem{Seiler:2012wz}
E.~Seiler, D.~Sexty, and I.-O. Stamatescu,
\newblock Phys.Lett. {\bf B723}, 213 (2013), arXiv:1211.3709.

\bibitem{Mollgaard:2014mga}
A.~Mollgaard and K.~Splittorff,
\newblock Phys.Rev. {\bf D91}, 036007 (2015), arXiv:1412.2729.

\bibitem{Mollgaard:2013qra}
A.~Mollgaard and K.~Splittorff,
\newblock Phys.Rev. {\bf D88}, 116007 (2013), arXiv:1309.4335.

\bibitem{Greensite:2014cxa}
J.~Greensite,
\newblock Phys. Rev. {\bf D90}, 114507 (2014), arXiv:1406.4558.

\bibitem{Nishimura:2015pba}
J.~Nishimura and S.~Shimasaki,
\newblock Phys. Rev. {\bf D92}, 011501 (2015), arXiv:1504.08359.

\bibitem{Pham:1983}
F.~Pham,
\newblock Proc. Symp. Pure Math. {\bf 40}, 319 (1983).

\bibitem{Witten:2010cx}
E.~Witten,
\newblock {Analytic Continuation Of Chern-Simons Theory},
\newblock in {\em {Chern-Simons gauge theory: 20 years after. Proceedings,
  Workshop, Bonn, Germany, August 3-7, 2009}}, pp. 347--446, 2010,
  arXiv:1001.2933.

\bibitem{Cristoforetti:2012su}
M.~Cristoforetti, F.~Di~Renzo, and L.~Scorzato,
\newblock Phys. Rev. {\bf D86}, 074506 (2012), arXiv:1205.3996.

\bibitem{Cristoforetti:2013wha}
M.~Cristoforetti, F.~Di~Renzo, A.~Mukherjee, and L.~Scorzato,
\newblock Phys. Rev. {\bf D88}, 051501 (2013), arXiv:1303.7204.

\bibitem{Fujii:2013sra}
H.~Fujii {\em et~al.},
\newblock JHEP {\bf 10}, 147 (2013), arXiv:1309.4371.

\bibitem{Mukherjee:2014hsa}
A.~Mukherjee and M.~Cristoforetti,
\newblock Phys. Rev. {\bf B90}, 035134 (2014), arXiv:1403.5680.

\bibitem{Tanizaki:2014tua}
Y.~Tanizaki,
\newblock Phys. Rev. {\bf D91}, 036002 (2015), arXiv:1412.1891.

\bibitem{DiRenzo:2015foa}
F.~Di~Renzo and G.~Eruzzi,
\newblock Phys. Rev. {\bf D92}, 085030 (2015), arXiv:1507.03858.

\bibitem{Fukushima:2015qza}
K.~Fukushima and Y.~Tanizaki,
\newblock PTEP {\bf 2015}, 111A01 (2015), arXiv:1507.07351.

\bibitem{Tanizaki:2015rda}
Y.~Tanizaki, Y.~Hidaka, and T.~Hayata,
\newblock New J. Phys. {\bf 18}, 033002 (2016), arXiv:1509.07146.

\bibitem{Fujii:2015bua}
H.~Fujii, S.~Kamata, and Y.~Kikukawa,
\newblock JHEP {\bf 11}, 078 (2015), arXiv:1509.08176,
\newblock [Erratum: JHEP02,036(2016)].

\bibitem{Fujii:2015vha}
H.~Fujii, S.~Kamata, and Y.~Kikukawa,
\newblock JHEP {\bf 12}, 125 (2015), arXiv:1509.09141.

\bibitem{Alexandru:2015xva}
A.~Alexandru, G.~Basar, and P.~Bedaque,
\newblock Phys. Rev. {\bf D93}, 014504 (2016), arXiv:1510.03258.

\bibitem{Kanazawa:2014qma}
T.~Kanazawa and Y.~Tanizaki,
\newblock JHEP {\bf 1503}, 044 (2015), arXiv:1412.2802.

\bibitem{Aarts:2013fpa}
G.~Aarts,
\newblock Phys. Rev. {\bf D88}, 094501 (2013), arXiv:1308.4811.

\bibitem{Aarts:2014nxa}
G.~Aarts, L.~Bongiovanni, E.~Seiler, and D.~Sexty,
\newblock JHEP {\bf 1410}, 159 (2014), arXiv:1407.2090.

\bibitem{Aarts:2013uza}
G.~Aarts, P.~Giudice, and E.~Seiler,
\newblock Annals Phys. {\bf 337}, 238 (2013), arXiv:1306.3075.

\bibitem{Hayata:2015lzj}
T.~Hayata, Y.~Hidaka, and Y.~Tanizaki,
\newblock Nucl. Phys. {\bf B911}, 94 (2016), arXiv:1511.02437.

\end{thebibliography}

\end{document}